\documentclass[12pt]{article}
\usepackage{graphicx}
\usepackage[backend = bibtex,
			sorting = none, 
			style=numeric-comp,
			url=false,
			isbn=false]{biblatex}
\bibliography{library}
\usepackage{authblk}
\usepackage{amsfonts}
\usepackage{amssymb}
\usepackage{amsmath}
\usepackage{graphicx}
\usepackage[left=2cm,right=2cm,top=2cm,bottom=2cm]{geometry}
\usepackage{subfig}
\usepackage[dvipsnames]{xcolor}
\usepackage{hyperref}
\usepackage{booktabs}

\newcommand\numberthis{\addtocounter{equation}{1}\tag{\theequation}}

\renewcommand{\d}{\mathrm{d}}
\newcommand{\iu}{\mathrm{i}}
\newcommand{\e}[1]{\mathrm{e}^{#1}}
\newcommand{\deltab}{\Delta\beta}
\newcommand{\moddeltab}{\overline{\Delta\beta}}
\newcommand{\sinc}[1]{\mathrm{sinc}\left({#1}\right)}

\title{General analytic theory of classical collinear three wave mixing in a monolithic cavity.}

\author[$\dagger$,*]{Matteo Santandrea}
\author[$\dagger$]{Michael Stefszky}
\author[$\dagger$]{Christine Silberhorn}
\affil[$\dagger$]{Integrated Quantum Optics, Paderborn University, Warburgerstr. 100, 33098 Paderborn, Germany}
\affil[*]{matteo.santandrea@upb.de}

\begin{document}
\maketitle


\begin{abstract}
Integrated, monolithic nonlinear cavities are of high interest in both classical and quantum optics experiments for their high efficiency and stability. 
However, a general, analytic theory of classical three wave mixing in such systems that encompasses multiple monolithic designs, including both linear and nonlinear regions, as well as any three-wave mixing process has not yet been fully developed. 

In this paper, we present the analytic theory for a general, classical three wave mixing process in a cavity with arbitrary finesse and non-zero propagation losses, encompassing second harmonic, sum frequency and difference frequency generation - SHG, SFG and DFG respectively. The analytic expression is derived under the sole assumption of low single-pass conversion efficiency (or equivalently operating in the non-depleted pump regime).

We demonstrate remarkable agreement between the presented model and the experimentally obtained highly complex second-harmonic spectrum of a titanium-diffused lithium niobate waveguide cavity that includes both a linear and nonlinear section. We then show the effect that reversing the linear and nolinear regions has on the output spectrum, highligthing the importance of system design. Finally, we demonstrate that the model can be extended to include the effect of phase modulation applied to the cavity.
\end{abstract}

%
%
%
%
%

\section{Introduction}
Integrated devices offer greater stability, easier interfacing to fiber networks and smaller footprint than their bulk counterparts \cite{Orieux2016}.
The functionality of these devices can be extended by incorporating a wide variety of linear (e.g. directional couplers, phase shifters) and nonlinear (e.g. second harmonic generation stages, polarisation converter) components on chip \cite{Lenzini2018, Luo2018}.
This flexibility makes integrated nonlinear optical devices key components of many classical and quantum optical experiments.

In recent years, integrated monolithic cavities have gained increasing interest, in particular in the quantum optics community \cite{Yonezawa:10, Phillips2011, PhysRevLett.108.153605, Brieussel:16, Zhang2017, Alsing2017, Yao2018}. They enable the generation of pure parametric downconverted states \cite{Luo2015}, of optical frequency combs \cite{Stefszky2018}, and squeezing \cite{Stefszky2017}, and can be used as interconnecting blocks for interfacing with narrowband quantum memories \cite{Zielinska2017}.

Such advantages comes with a price, namely an increased complexity of the integrated structure and less access to certain degrees of freedom to address this increased complexity. In particular, many recent nonlinear integrated cavities report the presence of a nonlinear section surrounded by electrooptic \cite{Stefszky2020} or thermooptic and piezooptic \cite{Zielinska2017} modulators. This allows a fine tuning of the resonance conditions but results in a more complex interaction between the three fields inside the cavity that cannot be addressed using more standard bulk dispersion compensation techniques, such as using a wedged crystal \cite{Imeshev1998, Stefszky2010}.

To date, the main method to study these systems is based on the model developed by Berger in Ref. \cite{Berger1997}. However, this model only considers type 0 second harmonic generation (SHG) in a cavity comprising only a single nonlinear region, and therefore it cannot be directly extended to the wider class of nonlinear integrated cavities described above. 
An improvement over Berger's model is found in \cite{Zielinska2017}, where the authors model a type 0 SHG cavity with thermooptic modulators surrounding the nonlinear region. 
However, the model derived cannot be easily extended to different system configuration (e.g. type II SHG or sum frequency generation). Moreover, they neglect the impact of losses for the fundamental field, a condition that is reasonable for their bulk system but cannot be generalised e.g. to waveguide resonators. 

In this paper, we expand the model of \cite{Berger1997} and \cite{Zielinska2017} and provide a comprehensive analytical model of three wave mixing processes in nonlinear cavities under the non-pump depletion approximation. This general model is valid for second harmonic generation (SHG), sum frequency generation (SFG) and difference frequency generation (DFG)  processes  and therefore can also be used to derive useful insights into parametric downconversion (PDC) in cavity. We derive the equation describing the spectrum of the light generated in the cavity through a chosen nonlinear process and show that our model accurately reproduces experimental measurements performed on titanium-diffused lithium niobate waveguides.

\section{Analytic theory}
\begin{figure}[hbtp]
\centering
\includegraphics[width = 0.8\textwidth]{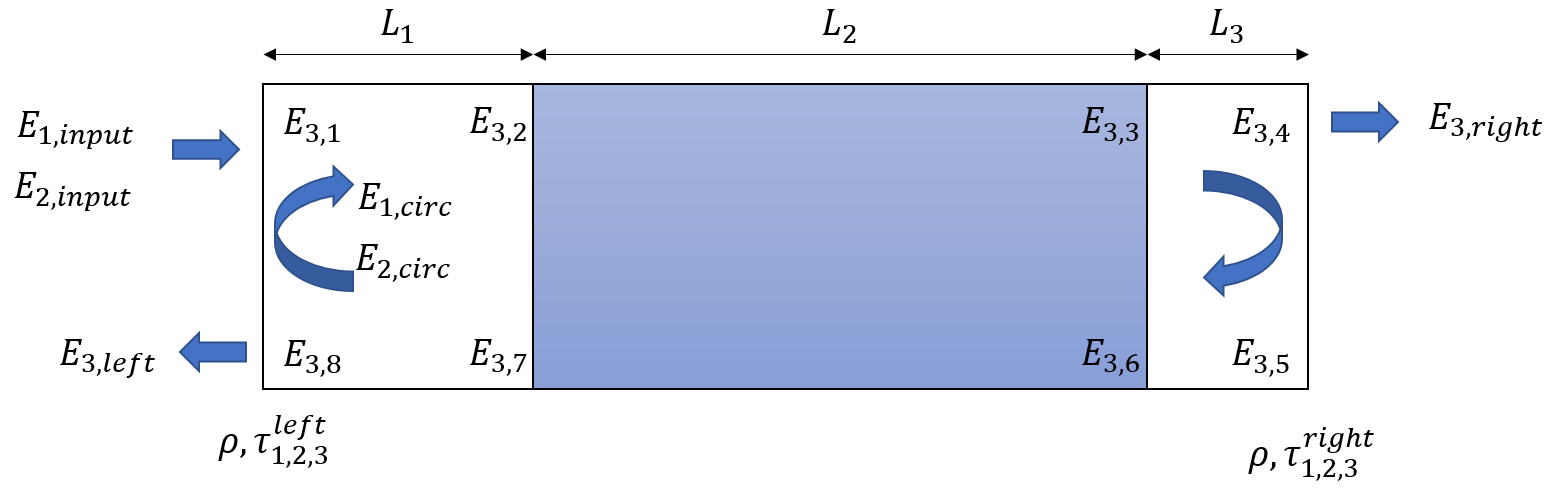}
\caption{Sketch of the system}
\label{img:sketch}
\end{figure}

Let us consider the system of length $L_{tot}=L_1+L_2+L_3$ sketched in Figure \ref{img:sketch}, composed of a nonlinear region of length $L_2$ surrounded by two regions without nonlinearity of length $L_1$ and $L_3$. 
Two input fields $E_1$ and $E_2$, at frequencies $\omega_1$ and $\omega_2$, enter the system from the left facet and generate the field $E_3$ at $\omega_3$ in the central nonlinear region. 
Due to non-zero facet (amplitude) reflection coefficients $\rho_{1,2,3}$ (with corresponding transmission coefficients $\tau_{1,2,3}$), the three fields interfere with themselves as they propagate back and forth through the sample. Therefore, the spectrum of the generated field $E_3$ will be the result of the phase-matching properties of the nonlinear region as well as the resonance conditions of the three fields, which we now aim to describe. 
An analytic solution for the spectrum $E_3(\omega_3)$ can be found under the assumption that the input fields $E_{1,2}$ are undepleted by the nonlinear process. Removing this assumption is possible, but the solution requires a numerically based iterative approach \cite{Fujimura1996}.

We start by considering the evolution of the fields $E_1$ and $E_2$, which enter the system from the left facet. Under the non-depletion approximation, the steady-state circulating fields inside the cavity, calculated at the left facet of the system, are given by the usual Fabry-P\'erot resonance condition
\begin{equation}
E_{j,\, circ} = \frac{\tau_j^{left}}{1-\rho_j^{left}\rho_j^{right}\e{-2 \iu \phi_{FP,j}}} E_{j, input}\qquad j=1,2
\end{equation}
where $\phi_{FP, j}$ is the complex phase factor acquired by the field $j$ over a single round trip of the cavity, which is given by 
\begin{align*}
    \phi_{FP,j} &= \phi_{j,1}+\phi_{j,2}+\phi_{j,3}\\
    \phi_{j,l} &= (\beta_{j,l} - \iu \alpha_j/2)L_l\qquad l=1,\,2,\,3.
\end{align*}

Here, $\beta_{j,l} = 2\pi n_{j,l}/\lambda_j$ is the propagation constant of field $j$ in region $l$ and $\alpha_j$ is the intensity propagation losses of field $j$. We consider the propagation constants in the three regions to be independent, allowing the description of systems with active elements that can modify the phase relationship between the fields, e.g. electro-optic or thermo-optic modulators.

The following step involves tracking the evolution of the field $E_3$ along a single round trip of the cavity. This leads to the following system of equations
\begin{equation}
\left\{
                \begin{array}{ll}
E_{3,2} &=  E_{3,1} \e{-\iu \phi_{3,1}}\\
E_{3,3} &=  NL(L_2; E_{1,2},\,E_{2,2},\,E_{3,2} )\e{-\iu \beta_{3,2} L_2}\\
E_{3,4} &=  E_{3,3}\e{-\iu \phi_{3,3}}\\
E_{3,5} &=  \rho_3^{right}E_{3,4}\\
E_{3,6} &=  E_{3,5}\e{-\iu \phi_{3,1}}\\
E_{3,7} &=  NL(L_2; E_{1,6},\,E_{2,6},\,E_{3,6} )\e{-\iu \beta_{3,2} L_2}\\
E_{3,8} &=  E_{3,7}\e{-\iu \phi_{3,1}}\\
E_{3,1} &=  \rho_3^{left}E_{3,8}              
                \end{array}
              \right.         
\label{eq:system_eq}
\end{equation}
with 
\begin{align*}
E_{1,2} &= E_{1,circ} \e{-\iu \phi_{1,1}}\\
E_{2,2} &= E_{2,circ} \e{-\iu \phi_{2,1}}\\
E_{1,6} &= E_{1,circ} \e{-\iu (\phi_{1,1}+\phi_{1,2}+2\phi_{1,3})}\\
E_{2,6} &= E_{2,circ} \e{-\iu (\phi_{2,1}+\phi_{2,2}+2\phi_{2,3})}.
\end{align*}

In Equation (\ref{eq:system_eq}), the notation $NL(L; E_{1},\,E_{2},\,E_{3} )$ has been used to indicate the complex amplitude of the field $E_3$ generated in a section $L$ of nonlinear material, with initial conditions $E_{1}$, $E_{2}$ and $E_{3}$.
In the case of sum frequency generation (SFG), where $\omega_3 =\omega_1 + \omega_2$, the evolution of the field $E_3$ can be evaluated solving the coupled system of equations
\begin{align*}
\frac{\d A_1}{\d z} &=  - \frac{\alpha_1}{2}A_1\\
\frac{\d A_2}{\d z} &=  - \frac{\alpha_2}{2}A_2\\
\frac{\d A_3}{\d z} &= \iu \gamma A_1(z) A_2(z)\e{\iu \deltab z} - \frac{\alpha_3}{2}A_3.\numberthis
\label{eq:coupled_system}
\end{align*}
In Eq. \eqref{eq:coupled_system}, $\Delta\beta = \beta_{3,2}-\beta_{2,2} - \beta_{1,2} -\beta_{G}$ is the wavevector mismatch, where $\beta_G$ is the additional quasi-phase matching grating vector, when necessary. 
The solution of Equation (\ref{eq:coupled_system}) over a nonlinear region of length $L$ and with initial conditions $A_{1,0}$, $A_{2,0}$, $A_{3,0}$ can be found analytically, resulting in 
\begin{align*}
A_3(L) &= NL_3(L; A_{1,0},\,A_{2,0},\,A_{3,0})\\
&=\iu \gamma A_{1,0} A_{2,0} L  \sinc{\frac{\moddeltab L}{2} }\e{\iu \frac{\moddeltab L}{2}}\e{-\frac{\alpha_3 L}{2}} +A_{3,0}\e{-\frac{\alpha_3 L}{2}},\numberthis
\label{eq:phasematching}
\end{align*}
where $\moddeltab$ is given by
\begin{equation}
\moddeltab = \Delta\beta - \iu (\alpha_3 - \alpha_2 -\alpha_1)/2.
\end{equation}

The self-consistent system \eqref{eq:system_eq} and the phase matching equation \eqref{eq:phasematching} allows one to retrieve the value of the field $E_3$ propagating inside the cavity. Finally, one can propagate the intra-cavity field $E_3$ to the external fields outside the cavity via the amplitude transmission coefficients $\tau_3$. The resulting electric fields can be factored as follows
\begin{align*}
E_3^{out, right} &=  \tau_3^{right} \tau_1^{left}\tau_2^{left}E_{1, input}E_{2, input}\cdot PM\cdot FP_1\cdot FP_2\cdot FP_3 \cdot \Phi_r\\
E_3^{out, left} &= \tau_3^{left} \tau_1^{left}\tau_2^{left}E_{1, input}E_{2, input} \cdot PM\cdot FP_1\cdot FP_2\cdot FP_3 \cdot \Phi_l\numberthis
\label{eq:general_solution}
\end{align*}
with 
\begin{align*}
PM &= \iu \gamma L_2\sinc{\frac{\moddeltab L_2}{2} }\e{\iu \frac{\moddeltab L_2}{2}} \\
FP_{j} &=  \frac{1}{1-\rho_{j}^{left}\rho_{j}^{right}\e{-2 \iu \phi_{FP,j}}},\qquad j = 1,2,3\\
\Phi_r &= \e{-\iu(\phi_{1,1} + \phi_{2,1} + \phi_{3,2} + \phi_{3,3}) } \times\\
&\qquad(1 + \rho_1^{right}\rho_2^{right} \rho_3^{left} \e{-\iu (2\phi_{3,1}+\phi_{1,2}+ \phi_{2,2} + \phi_{3,2} + 2\phi_{1,3} +2 \phi_{2,3}  )})\\
\Phi_l &= \e{-\iu ( \phi_{1,1} +\phi_{2,1} + \phi_{3,1} + \phi_{1,2} +\phi_{2,2} -2\phi_{FP,1} -2\phi_{FP,2} - 2\phi_{FP,3}))}\times\\&
\left( \rho_1^{right}\rho_2^{right}\e{\iu (\phi_{3,2}+ 2\phi_{3,3})} + \rho_3^{right}\e{\iu (\phi_{1,2}+\phi_{2,2}+2\phi_{1,3}+2\phi_{2,3})}\right)\numberthis
\label{eq:general_factors}
\end{align*}
In writing Eq. \ref{eq:general_solution}, the different factors determining the spectrum of $E_3$ outside the cavity have been separated according to their source: the term $PM$ represents the (single-pass) phase-matching spectrum of the nonlinear section $L_2$, the terms $FP_{1/2/3}$ represent the contributions of the Fabry-P\'erot cavities of the three fields and the terms $\Phi_{r/l}$ represent the interference between the forward and backward generated $E_3$ fields.

Equations \eqref{eq:general_solution} and \eqref{eq:general_factors} represent the main result of our work. 
In the next section, we will show how this model accurately models the response of a real cavity system.


\section{Application to a real system}

Here, we model the spectral properties of the double pass second harmonic cavity system presented in \cite{Stefszky2020}. 
The device consists of a $\sim$2cm-long waveguide comprising a $\sim$1cm-long unpoled region with an electrooptical phase modulator on the left (input) side followed by a $\sim$1cm-long poled region for second harmonic generation pumped at 1540nm on the right (output) side. 

Dielectric coatings are deposited on the waveguide facets, such that the waveguide acts as a high finesse resonator for the fundamental field and a double-pass structure for the second harmonic field.
The input facet is chosen to have a 70\% (intensity) reflectivity for the fundamental field and a high reflectivity coating for the second harmonic, while the output facet has a high reflectivity coating for the fundamental field and an anti-reflection coating for the second harmonic field.
The measured values for the reflection coefficients $\rho$ and the propagation losses $\alpha_{FF}$ for the fundamental field are summarised in Table \ref{table:measured_parameters}. The losses $\alpha_{SH}$ of the second harmonic field are difficult to characterise as the waveguide is multimode at this frequency and thus are assumed to be twice as high as the fundamental losses. We also note that, owing to the double-pass configuration, the second harmonic losses will only have a minor impact on the output spectrum of the device.

\begin{table}
\centering
\begin{tabular}{ cccccc } 
 \toprule
 $\rho_{FF}^{left}$ & $\rho_{FF}^{right}$ & $\rho_{SH}^{left}$ & $\rho_{SH}^{right}$ & $\alpha_{FF}$   & $\alpha_{SH}$ \\ 
 \midrule
 0.833 & 0.998 & 0.998 & 0.1 & 0.25 dB cm$^{-1}$ & 0.5 dB cm$^{-1}$\\ 
 \bottomrule
\end{tabular}
\caption{Values of the reflection coefficients and fundamental losses used to model the system presented in \cite{Stefszky2020}.}
\label{table:measured_parameters}
\end{table}

\begin{figure}
    \centering
    \includegraphics[width = 0.6\textwidth]{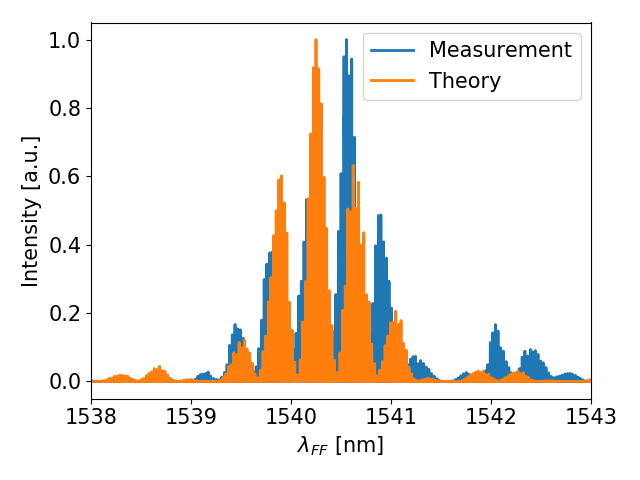}
    \caption{Measured phase matching spectrum of the double pass SHG device reported in Ref. \cite{Stefszky2020} and its theoretical model found using Eq. \ref{eq:general_solution}.}
    \label{fig:modelling_doublepass}
\end{figure}

Under these assumptions, it is possible to calculate the phase matching spectrum of the waveguide cavity using Eqs. \eqref{eq:general_solution} and \eqref{eq:general_factors}. 
The comparison between the measured phase matching spectrum and the theoretical model is shown in Figure \eqref{fig:modelling_doublepass}. Our model is able to capture remarkably well most of the characteristics of the highly complex measured spectrum. The discrepancies between the theoretical and the measured spectra arise from the difficulty in estimating the exact length of each section and from the inhomogeneities present in the sample, further detailed in \cite{Stefszky2020}.

\begin{figure}
    \centering
    \includegraphics[width = 0.64\textwidth]{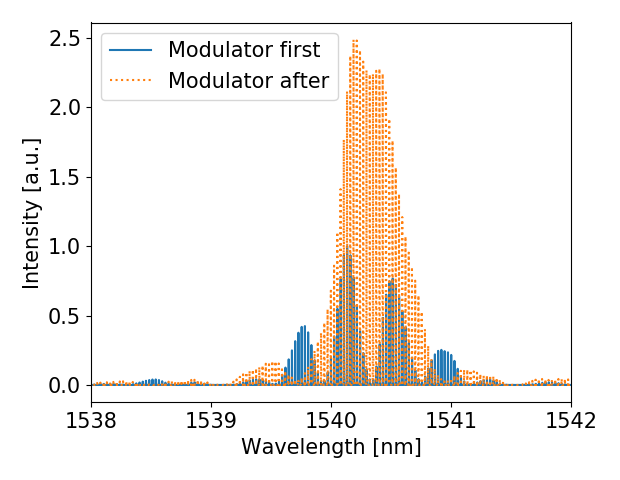}
    \caption{Expected phase matching spectra for the sample presented in \cite{Stefszky2020} and modelled with the parameters reported in Table \ref{table:measured_parameters}, depending on the position of the phase modulator with respect to the nonlinear region.}
    \label{fig:comparison_orientation}
\end{figure}

The theory developed in the previous section allows one to investigate the impact that system design has on device performance. As an example, we investigate the impact that swapping the order of the phase modulator and nonlinear region has on the output spectrum. The resulting phase matching spectra are shown in Figure \ref{fig:comparison_orientation}.
One can see that the spectrum of the sample with the phase modulator before the nonlinear region (as is the case for the system presented in \cite{Stefszky2020}) is characterised by a periodic modulation of the envelope, which is absent in the opposite configuration. Moreover, its peak intensity is $\sim$20\% of the maximum intensity generated by the sample with the modulator placed after the nonlinear region.
This reveals that the resonance conditions in the first configuration are much stricter and, depending on the application, may impact device performance. This highlights the importance of careful design and modeling of these systems.

Finally, this model also allows one to calculate the effect that driving the phase modulator has on the output phase matching spectrum. This is achieved by writing the refractive index of the fields in the first region, i.e. where the electrooptic modulator is located, as 
\begin{equation}
    n(V) = n_e - \frac{n_e^3}{2}r \frac{V}{d}\Gamma
\end{equation}
where $r$ is the electrooptic coefficient addressed by the phase modulator (in this case, it corresponds to $r_{33} = 30.8$pm/V), $V$ is the voltage applied to the electrodes of the modulator, $d$ is the distance between the electrodes, $n$ is the extraordinary refractive index of the waveguide in the absence of any electric field and $\Gamma$ is the overlap integral between the field of the modulator and the one of the guided mode.

\begin{figure}
    \centering
    \includegraphics[width = 0.6\textwidth]{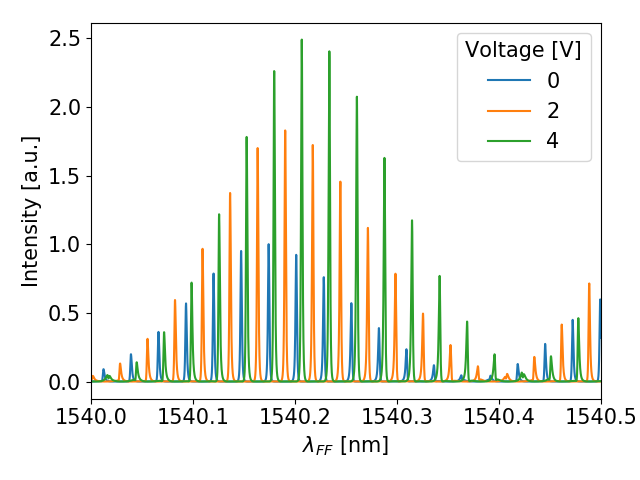}
    \caption{Predicted shift of the phase matching spectrum with various voltages applied to the electrooptic modulator for the sample presented in \cite{Stefszky2020} and modelled with the parameters reported in Table \ref{table:measured_parameters}.}
    \label{fig:modulator}
\end{figure}

As a non-zero voltage is applied to the modulator, the phase of both the fundamental and second harmonic field varies, thus changing the interference conditions inside the cavity. This results in a shift in the position of the resonance peaks of the structure, as illustrated in Figure \ref{fig:modulator} for three different voltage levels. 

\section{Conclusions}
In this paper we presented a general, analytic theory of an integrated cavity comprising linear and nonlinear sections under the sole approximation of no pump depletion. 
The presented model is general and can be applied to any type of collinear three wave mixing process - both in bulk and in waveguides. We demonstrated that it accurately reproduces experimental data and demonstrated its capabilities in modelling the effect of integrated modulators on the phase matching spectrum of the device. 
This model constitutes a fundamental step towards the understanding and optimization of the performance of a wide variety of new, complex resonant nonlinear devices for both classical and quantum optics applications.


\printbibliography
\end{document}